UNIVERSIDADE FEDERAL DO PARANÁ

EDUARDO AUGUSTO ROEDER

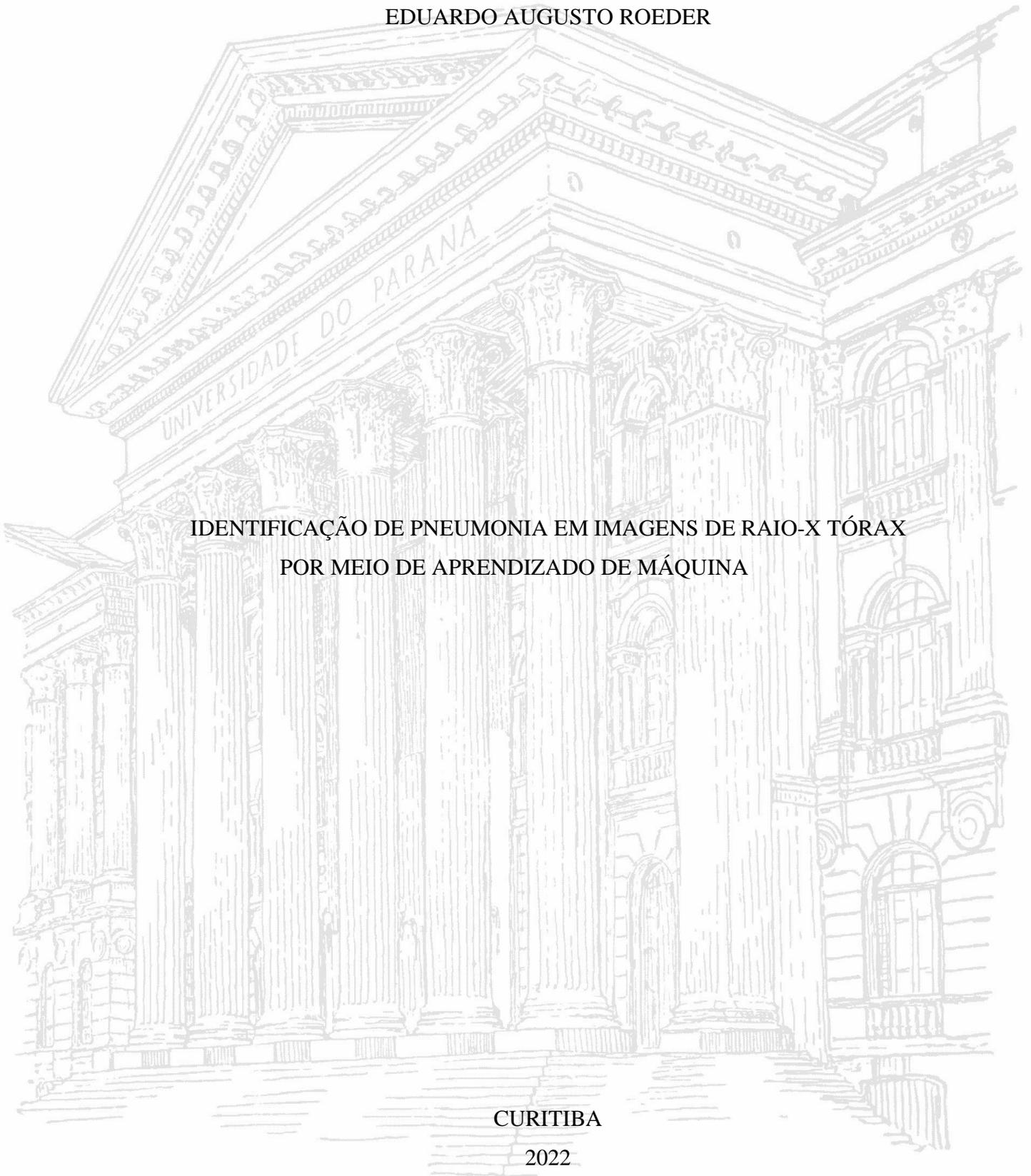

IDENTIFICAÇÃO DE PNEUMONIA EM IMAGENS DE RAIO-X TÓRAX
POR MEIO DE APRENDIZADO DE MÁQUINA

CURITIBA
2022

EDUARDO AUGUSTO ROEDER

IDENTIFICAÇÃO DE PNEUMONIA EM IMAGENS DE RAIO-X TÓRAX
POR MEIO DE APRENDIZADO DE MÁQUINA

Trabalho de Curso apresentado como requisito parcial
à finalização do Curso de Graduação em Medicina,
Complexo Hospital de Clínicas da Universidade
Federal do Paraná.

Orientador: Profº. Dr. Akihito Inca Atahualpa Urdiales

CURITIBA
2022

# TERMO DE APROVAÇÃO

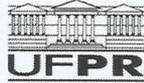

MINISTÉRIO DA EDUCAÇÃO
UNIVERSIDADE FEDERAL DO PARANÁ
SETOR DE CIÊNCIAS DA SAÚDE
COORDENAÇÃO DO CURSO DE MEDICINA

## DECLARAÇÃO

Eu, Profº. Dr. Akihito Inca Atahualpa Urdiales, declaro conhecer e aprovar o conteúdo do trabalho intitulado "IDENTIFICAÇÃO DE PNEUMONIA EM IMAGENS DE RAIO-X TÓRAX POR MEIO DE APRENDIZADO DE MÁQUINA" de autoria do(a) estudante Eduardo Augusto Roeder, GRR20159337, liberando-o para ser apresentado.

Estou também ciente da necessidade de minha participação na banca examinadora por ocasião da defesa do trabalho.

Curitiba, 12 de Maio de 2022

Assinatura e Carimbo

CRM 15873
CPF 859 360 228-00

*Dedico este trabalho aos meus pais e irmãos, pessoas que foram os alicerces da trajetória que me trouxeram até a realização deste sonho.*

# AGRADECIMENTOS

Em primeiro lugar, agradeço aos meus pais, Décio e Silmara, por estarem sempre presentes ainda que distantes. Sou grato também aos meus irmãos, Décio e Geovanna, por me apoiarem em momentos difíceis.

Aos meus avós maternos Juca (in memoriam) e Célia; aos meus avós paternos Nivaldo (in memoriam) e Dirce, que me olham de algum lugar.

Agradeço ainda a todos os professores que participaram da minha jornada acadêmica, em especial ao Prof. Dr. Akihito I. A. Urdiales, responsável pela orientação do meu projeto. Obrigado pela orientação, paciência e pela confiança.

Agradeço à UFPR e ao HC, por proporcionar um ambiente propício para o treinamento e aprendizado. Sou grato a todos os membros do corpo docente, à direção e a administração destas instituições.

Por fim, agradeço à minha banca examinadora, por destinar um tempo para avaliação do presente projeto.

*Por Eduardo Augusto Roeder*

# RESUMO

**INTRODUÇÃO:** Pneumonia é a principal causa infecciosa de morte infantil no mundo. Quando identificada precocemente, é possível alterar o prognóstico do paciente, podendo ser utilizado exame de imagem para confirmação diagnóstica. Executar e interpretar os exames o quanto antes é vital para um bom tratamento, sendo o exame mais comum para este tipo de patologia o raio-X torácico. O objetivo deste trabalho foi desenvolver um software que identifique a presença ou ausência de pneumonia em radiografias de tórax. **MATERIAL E MÉTODOS**: O software foi desenvolvido como um modelo computacional através de aprendizado de máquina utilizando a técnica de transferência de aprendizado. Para o treinamento, imagens foram coletadas de um banco de dados disponível online com imagens de raio-X torácico infantil realizadas em um hospital na China. **RESULTADOS:** Após o treinamento o modelo foi exposto a imagens novas, atingindo resultados relevantes para identificação de tal patologia com 98% de sensibilidade e 97.3% de especificidade na amostra utilizada para teste. **CONCLUSÃO**: Pode-se concluir que é possível desenvolver um software que identifique pneumonia em imagens de raio-X torácico.

Palavras-chave: pneumonia; radiografia; raio-X torácico; aprendizado de máquina.

**RESUMO**

Pneumonia é a principal causa infecciosa de morte infantil no mundo. Quando identificada precocemente, é possível alterar o prognóstico do paciente, podendo ser utilizado exame de imagem para confirmação diagnóstica. Executar e interpretar os exames o quanto antes é vital para um bom tratamento, sendo o exame mais comum para este tipo de patologia o raio-X torácico. O objetivo deste trabalho foi desenvolver um software que identifique a presença ou ausência de pneumonia em radiografias de tórax. O software foi desenvolvido como um modelo computacional através de aprendizado de máquina utilizando a técnica de transferência de aprendizado. Para o treinamento, imagens foram coletadas de um banco de dados disponível online com imagens de raio-X torácico infantil realizadas em um hospital na China. Após o treinamento o modelo foi exposto a imagens novas, atingindo resultados relevantes para identificação de tal patologia com 98% de sensibilidade e 97.3% de especificidade na amostra utilizada para teste. Pode-se concluir que é possível desenvolver um software que identifique pneumonia em imagens de raio-X torácico.

Palavras-chave: pneumonia; radiografia; raio-X torácico; aprendizado de máquina.

## ***ABSTRACT***

Pneumonia is the leading infectious cause of infant death in the world. When identified early, it is possible to alter the prognosis of the patient, one could use imaging exams to help in the diagnostic confirmation. Performing and interpreting the exams as soon as possible is vital for a good treatment, with the most common exam for this pathology being chest X-ray. The objective of this study was to develop a software that identify the presence or absence of pneumonia in chest radiographs. The software was developed as a computational model based on machine learning using transfer learning technique. For the training process, images were collected from a database available online with children's chest X-rays images taken at a hospital in China. After training, the model was then exposed to new images, achieving relevant results on identifying such pathology, reaching 98% sensitivity and 97.3% specificity for the sample used for testing. It can be concluded that it is possible to develop a software that identifies pneumonia in chest X-ray images.

# LISTA DE ILUSTRAÇÕES

# LISTA DE TABELAS

# SUMÁRIO

# 1    INTRODUÇÃO

Segundo a Organização Mundial da Saúde (OMS), mortes por infecções de vias aéreas inferiores são a quarta maior causa de morte do mundo, sendo pneumonia a principal causa infecciosa de morte infantil no mundo (OMS, 2021), matando mais crianças do que HIV/AIDS, malária e sarampo juntas (ADEGBOLA, 2012). As recomendações para investigação compreendem realizar uma boa anamnese e um bom exame físico, com a clínica indicando sinais para uma melhor investigação com exames de imagem (CORRÊA *et al.*, 2018). O exame complementar inicial é a radiografia torácica, onde é possível identificar opacidades em focos pulmonares, permitindo avaliar extensões das lesões, detectar complicações e auxiliar no diagnóstico diferencial (CORRÊA *et al.*, 2018). Embora a pneumonia possa ter diferentes etiologias, na maioria das vezes é impossível diferenciá-las apenas com radiografias (MARRIE, 1994). Considerando a gravidade de uma pneumonia, é vital que médicos identifiquem o quadro o quanto antes para que o tratamento seja iniciado o mais precocemente possível. Há estudos que mostram que médicos que atuam no pronto socorro e radiologistas discordam com frequência dos resultados de radiografias em relação à pneumonia (CAMPBELL *et al.*, 2005; ATAMNA *et al.*, 2019), o que pode atrasar a confirmação diagnóstica.

Exames de imagem são vitais para investigação e confirmação diagnóstica de diversas patologias, com frequência direcionando diretamente a conduta médica. Dentre as opções de diagnóstico por imagem, o raio-X (RX) é o mais comum (NHS ENGLAND, 2021; BEKAS *et al.*, 2013) com cerca de 3.6 bilhões de radiografias executadas todos os anos (ORGANIZAÇÃO PANAMERICANA DE SAÚDE, 2012). A distribuição da execução desses exames varia com o nível de desenvolvimento do país, onde países menos desenvolvidos tendem a ter menos acesso a essas tecnologias quando comparados a países mais ricos, com cerca de dois terços da população mundial sem acesso a nenhum tipo de diagnóstico por imagem (ORGANIZAÇÃO PANAMERICANA DE SAÚDE, 2012).

Juntamente com a carência da tecnologia em países menos desenvolvidos, há espaço para que profissionais ainda que qualificados possam estar sujeitos ao estresse e sobrecarga e não percebam detalhes por vezes importantes, seja por desatenção ou por não ter o treinamento adequado, gerando laudos imprecisos ou incompletos (BASHIR *et al.*, 2019; FAWOLE, 2020; GONÇALVES, 2009). Laudos incorretos acarretam em prejuízos sistêmicos como a necessidade de uma segunda execução do exame ou de reavaliação do exame, aumentando a demora para confirmar o diagnóstico e causando assim um custo maior - tanto pela reexecução do exame quanto para os insumos utilizados pelo paciente nesse ínterim (GONÇALVES, 2009). Para remediar essas situações, há a possibilidade de solicitar uma segunda opinião, a qual também causa um gasto extra de recursos e requer ainda mais tempo (KHAN *et al.*, 2019). Uma alternativa à segunda opinião é automatizar a interpretação dessas imagens por meio de modelos computacionais.

Aprendizado de máquina, também chamado de *Machine Learning (ML),* consiste em modelos computacionais com objetivo de classificar informações novas de acordo com informações anteriormente apresentadas à máquina (CURRIE *et al.*, 2019). Esses modelos são treinados através de informações processadas e categorizadas com relação ao objetivo a ser atingido, adquirindo a capacidade de correlacionar essas informações novas dentro dos padrões encontrados durante o treinamento. Após o treinamento, o modelo é apresentado a dados novos, sugerindo a classificação desse dado relacionado aos dados utilizados em treinamento, exigindo uma baixa demanda de poder computacional podendo ser realizada em dispositivos mais simples, como smartphones. De regra geral, ocorre o efeito "caixa preta" pelo modelo não conseguir demonstrar diretamente o raciocínio por trás de sua classificação (LONDON, 2019), ainda que existam técnicas que tentem mostrar quais características o modelo utilizou para chegar uma conclusão.

## 2    OBJETIVO GERAL

Desenvolver um software que identifique a presença ou a ausência de pneumonia em imagens de raio-X torácico.

## 2.1 OBJETIVOS ESPECÍFICOS

A) Obter imagens de raio-X de tórax em bancos de dados públicos, já classificadas com a presença ou ausência de pneumonia;

B) Analisar e comparar os grupos das imagens coletadas;

C) Separar as imagens em três grupos, sendo um grupo para treino, um para teste e outro para validação;

D) Desenvolver um modelo de rede neural apto para receber e interpretar as imagens classificadas;

E) Treinar o modelo desenvolvido com o conjunto de imagens para treino;

F) Utilizar o modelo treinado para classificar as imagens do grupo de teste e avaliar sua performance.

G) Aplicar a técnica Grad-CAM para que o resultado do modelo seja transparente ao observador.

# 3    MATERIAIS E MÉTODOS

## 3.1    ASPECTOS ÉTICOS DA PESQUISA

Não foi necessário aprovação em Comitê de Ética visto que se tratam de dados públicos disponíveis em diferentes plataformas online.

## 3.2    DADOS UTILIZADOS

Foi utilizado um banco de imagens, que consiste em uma coleção de imagens catalogadas com informações relevantes disponíveis publicamente na internet. O banco de imagens utilizado é composto por 5856 imagens provenientes de exames raio-X anteroposterior (AP), de crianças com idade entre 1 a 5 anos, coletados em exames de rotina no Centro Médico de Mulheres e Crianças de Guangzhou (GWCMC) em Guangzhou, China e disponibilizado por Kermany *et al.* (2018). As imagens foram inicialmente interpretadas por três profissionais experientes, catalogando a presença ou ausência de pneumonia, sendo 1583 imagens classificadas como normais e 4253 classificadas como pneumonia. Foram salvas no formato JPEG e possuem dimensões variáveis, como apresentado na TABELA 1. A FIGURA 1 e a FIGURA 2 representam exemplos de imagens presentes em cada grupo. Para análise dos dados utilizados, os grupos de imagens foram comparados entre si, gerando uma imagem que representa a média de cada categoria.

TABELA 1 – DISTRIBUIÇÃO DAS IMAGENS

| Categoria | Quantidade | Dimensões |
|:---:|:---:|:---:|
| Normal | 1583 | [496-2713] px x [912-2916] px |
| Pneumonia | 4253 | [127-2304] px x [384-2772] px |

FONTE: Adaptada de KERMANY *et al.* (2018).

FIGURA 1 – EXEMPLOS DE IMAGENS NORMAIS

FIGURA 2 – EXEMPLOS DE IMAGENS COM PNEUMONIA

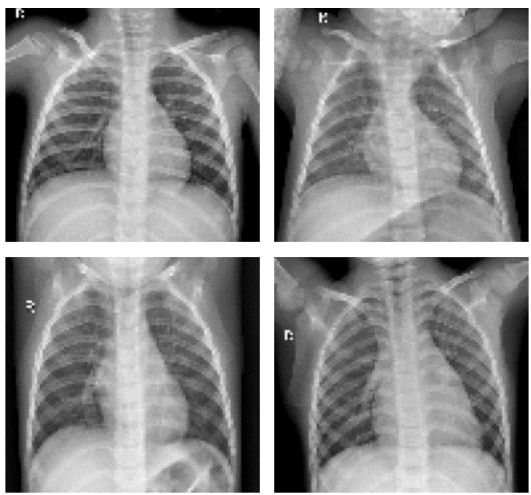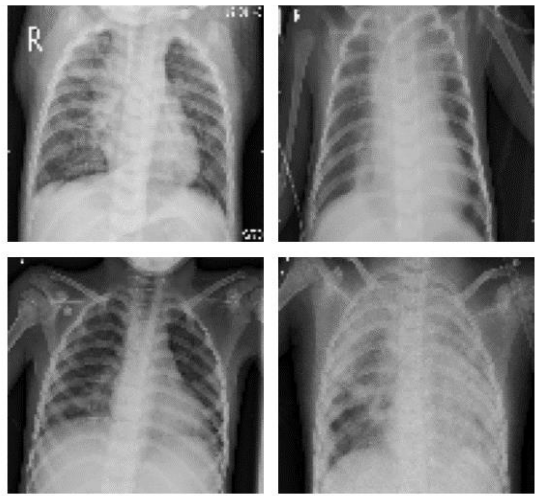

FONTE: Adaptado KERMANY *et al.* (2018).

## 3.3 PRÉ-PROCESSAMENTO DAS IMAGENS

As imagens foram redimensionadas a partir de sua resolução inicial para a resolução de 224 pixels (px) de largura por 224 pixels de altura. Em seguida, foram divididas em três grupos: um primeiro grupo para teste, um segundo grupo para treinamento e um terceiro grupo para validação. O grupo teste consiste em 300 imagens, sendo 150 imagens aleatoriamente selecionadas de cada categoria principal. O restante das imagens foi separado também de forma aleatória em uma proporção aproximada de 80% para treinamento e 20% para validação, ficando 4444 imagens para treinamento e 1112 imagens para validação. O grupo de treinamento foi submetido a um processo chamado *Image Augmentation* (PEREZ, 2017)**,** o qual consiste em submeter as imagens a transformações como rotação, distorção, deslocamento e zoom, o que adiciona uma variabilidade maior ao treinamento posterior (FIGURA 3).

TABELA 2 – SEPARAÇÃO DOS GRUPOS DE IMAGENS

| Categoria | Quantidade | Dimensões |
|:---:|:---:|:---:|
| Teste | 300 | |
| Treinamento | 4444 | 224px x 224px |
| Validação | 1112 | |

FONTE: O autor (2022).

FIGURA 3 – EXEMPLOS DE TRANSFORMAÇÕES APLICADAS

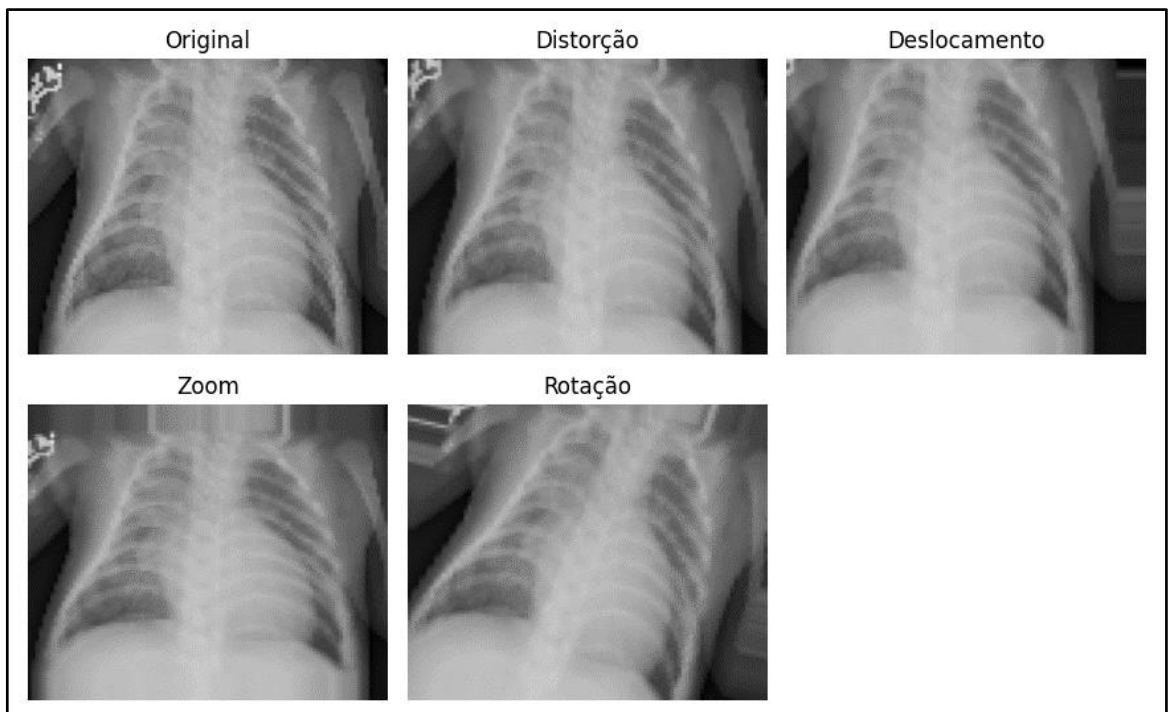

FONTE: O autor (2022).

## 3.4 DESENVOLVIMENTO

O software foi desenvolvido utilizando a linguagem de programação Python v3.9.7. Esta linguagem permite a utilização de conjuntos de código chamados de bibliotecas, as quais contém ferramentas para otimizar o desenvolvimento em geral. As bibliotecas utilizadas foram Keras (CHOLLET, 2015), TensorFlow (ABADI, 2016) e matplotlib (BARRET, 2005). Keras e TensorFlow são bibliotecas que otimizam e facilitam o desenvolvimento de modelos para aprendizado de máquina. Matplotlib fornece meios para plotagem de dados, que facilita a visualização das informações referentes ao treino e análise estatística.

### 3.4.1 Modelo de aprendizado de máquina

Softwares de aprendizado de máquina (ML) são modelos treinados para reconhecer certos padrões. Para este trabalho foi utilizado um modelo de ML, cuja estrutura consiste em redes neurais artificiais, as quais são compostas por camadas de forma sequencial (FIGURA 4). Cada camada possui uma quantidade arbitrária de neurônios, que por sua vez se relacionam diretamente com os neurônios das camadas adjacentes. Neste cenário, entende-se por neurônio uma estrutura que armazena os valores e pesos das relações que possuem com os neurônios das camadas adjacentes. Quando um modelo é treinado, essas relações são reajustadas e cada neurônio passa a adotar um valor diferente do anterior, ajustando assim os pesos de suas conexões.

FIGURA 4 – EXEMPLO DE REDE NEURAL

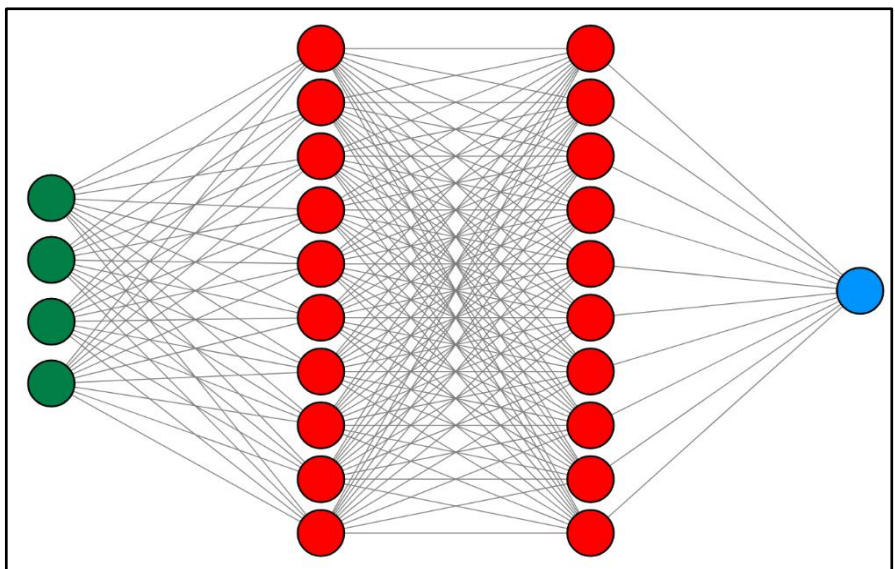

FONTE: O autor (2022).

Para fins de exemplo, a FIGURA 4 representa uma rede neural de 4 camadas. A primeira camada em verde, com 4 neurônios, é chamada de *input layer*, que corresponde a camada que recebe as informações externas para treinamento ou classificação. Em vermelho, há duas camadas intermediárias, cada uma com 10 neurônios. As camadas intermediárias são conhecidas como *hidden layers*. Em azul a camada de saída (*output layer*) está representada com apenas um neurônio. Por fim, em cinza, estão representadas as relações entre os neurônios em cada camada. Cada neurônio armazena um valor referente ao peso de cada conexão em que está envolvido, sendo este valor o que define o aprendizado de máquina em sua essência. Somando todas as relações desta imagem, há cerca de 290 parâmetros em uma rede como a exemplificada.

Para o presente trabalho foi utilizado a técnica de "transferência de aprendizado" (LU, 2015), a qual consiste em usar como base um modelo já treinado para um banco de dados prévio, adicionar camadas após o modelo base e treinar estas camadas finais para os dados em que desejamos classificar. Esta técnica permite aproveitar a performance de modelos com capacidade de extrair características de dados sem precisar treinar um novo modelo do início. O modelo utilizado como base foi o VGG19 (SIMONYAN, 2015) previamente treinado para o banco de imagens ImageNet (DENG, 2009) e disponibilizado pela biblioteca Keras. O VGG19 é uma rede neural de 19 camadas totalmente conectadas. Por ser previamente treinado em um banco de imagens com mais de 14 milhões de imagens, este modelo possui a capacidade de extrair características de imagens de forma otimizada. Em seguida foram adicionadas 4 camadas treináveis que aceitam o resultado do modelo base e retornam indicando se há a presença ou ausência de pneumonia na imagem fornecida. O modelo final, com 23 camadas, possui cerca de 46 milhões de parâmetros, sendo treináveis apenas aproximadamente 26 milhões. Os outros 20 milhões são parâmetros fixos vindos do modelo VGG19.

A estrutura do modelo proposto foi escolhida pelo pesquisador de forma arbitrária e de acordo com testes de diferentes variações. A estrutura final foi a que obteve melhores resultados estatísticos dentre as variações testadas.

## 3.4.2  Treinamento

O modelo foi submetido ao processo de treinamento com imagens dos grupos de treinamento e de validação em uma máquina com o processador Intel Core i7-7700k, 32GB de memória RAM e placa de vídeo NVidia GeForce 1080ti com 11GB de VRAM. O treinamento foi realizado por um total de 150 épocas, durando cerca de 3 horas no total. Uma época é o período em que o modelo é exposto para todos os dados do grupo de treinamento. Dentro de uma época, os dados foram divididos em lotes de 32 imagens, de forma que a cada lote os pesos e valores dos neurônios sejam ajustados em relação ao acerto ou erro na predição deste lote. Ao final de cada época, é medido o grau de acerto do modelo recém treinado relativo aos dados de validação para que se possa acompanhar sua performance em dados que o modelo não utilizou para treinamento.

## 3.4.3  Modelo de visualização

Para visualização das características que o modelo utilizou para chegar a uma conclusão, a técnica Grad-CAM (SELVARAJU, 2019) foi aplicada no modelo treinado. Esta técnica calcula dentro de uma imagem qualquer quais neurônios estão mais ativos e assim em quais pixels o modelo está "prestando mais atenção". Em seguida, é gerado um mapa de calor destas regiões, que em seguida é sobreposta à imagem original, o que permite visualizar quais são as regiões de interesse da imagem fornecida para o processo de avaliação pelo modelo, com objetivo de reduzir o efeito "caixa preta". Na FIGURA 5 e na FIGURA 6 estão ilustradas o mecanismo em uma imagem normal e em uma imagem com pneumonia, respectivamente. As áreas em roxo são regiões em que o modelo considerou com baixa importância para a predição, enquanto as áreas em verde ou amarelo são as regiões em que o modelo considerou com média e alta importância, respectivamente. Este processo não é necessário para a predição do modelo, porém permite maior clareza e transparência para o operador.

FIGURA 5 – GRAD-CAM APLICADA A UMA IMAGEM NORMAL

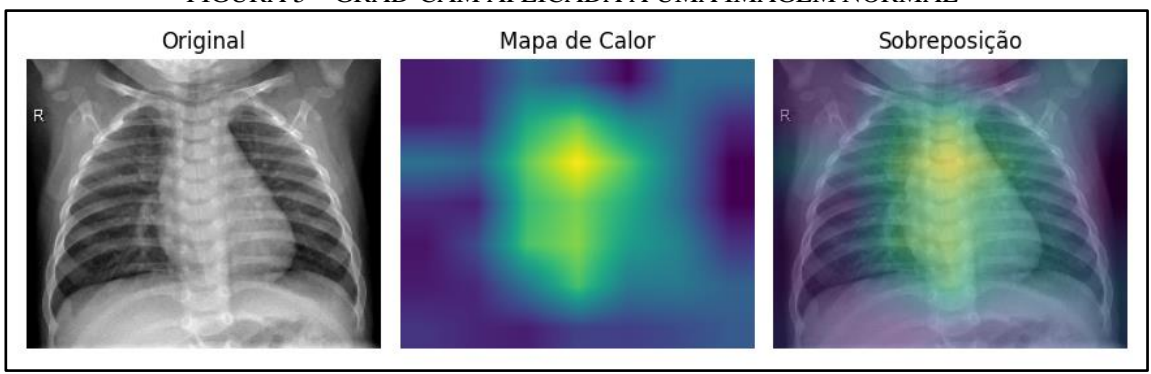

FIGURA 6 – GRAD-CAM APLICADA A UMA IMAGEM COM PNEUMONIA

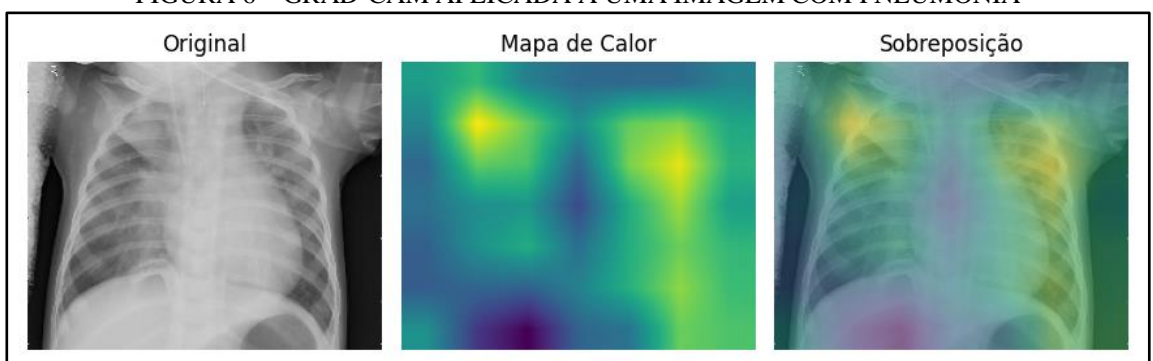

FONTE: O autor (2022).

## 3.5    ANÁLISE DOS DADOS

A análise dos dados foi realizada através da performance obtida ao longo do treinamento pelos grupos de treino e validação, e após o treinamento pelo grupo de teste. Quanto à performance do modelo em relação ao grupo de teste, foi realizado uma matriz de comparação de acordo com os valores obtidos como verdadeiro positivo (VP), verdadeiro negativo (VN), falso positivo (FP) e falso negativo (FN). Com estes dados foi calculado o valor preditivo positivo (VPP), valor preditivo negativo (VPN), especificidade (ESP), sensibilidade (SENS), acurácia (ACC) e o escore F1 para o modelo de acordo com as fórmulas presentes na FIGURA 7. Todas as imagens de análise foram geradas pela biblioteca matplotlib.

FIGURA 7 – FÓRMULAS UTILIZADAS PARA ANÁLISE

$$VPP = \frac{VP}{VP + FP} \qquad\qquad SENS = \frac{VP}{VP + FN}$$

$$VPN = \frac{FN}{FN + VN} \qquad\qquad F1 = 2 * \frac{ESP * SENS}{ESP + SENS}$$

$$ESP = \frac{VN}{VN + FP} \qquad\qquad ACC = \frac{VP * VN}{VP + VN + FN + FP}$$

FONTE: O autor (2022).

# 4 RESULTADOS

## 4.1 ANÁLISE DAS IMAGENS

As imagens coletadas foram analisadas calculando a imagem média para cada grupo. Para o grupo de imagens normais (FIGURA 9), a região pulmonar está mais escura, menos opaca. Para o grupo de imagens com pneumonia (FIGURA 8), a região pulmonar está com opacidade aumentada. A diferença entre os dois grupos denota a realidade ao interpretar um exame de raio-X. O mesmo é visto pela imagem gerada pela diferença entre as imagens médias (FIGURA 10), onde a cor azul mostra o que é mais comum no grupo das imagens normais; em vermelho é apresentado o que está predominante no grupo das imagens com pneumonia; enquanto na cor branca há um intermediário do que é comum a ambos os grupos.

FIGURA 9 – MÉDIA DAS IMAGENS NORMAIS

FIGURA 8 – MÉDIA DAS IMAGENS COM PNEUMONIA

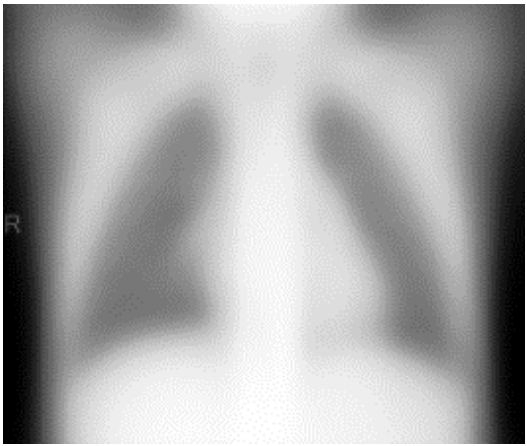 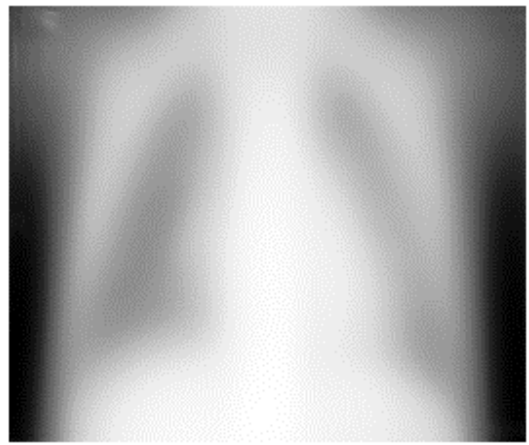

FIGURA 10 – DIFERENÇA ENTRE AS MÉDIAS DAS IMAGENS

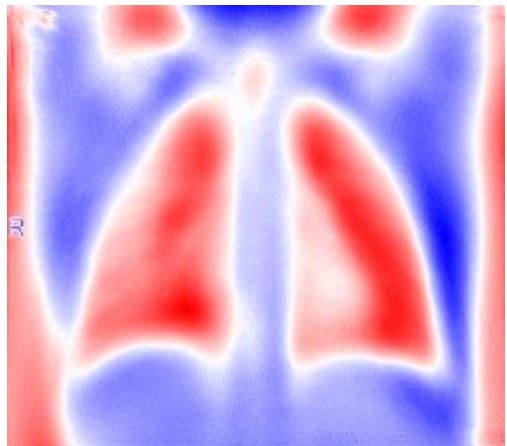

FONTE: O autor (2022).

## 4.2   TREINAMENTO

Durante as 150 épocas de treinamento do modelo proposto, a evolução da acurácia para o grupo de treino e para o grupo de validação pode ser observada na FIGURA 11. Esta variável é importante para acompanhamento do treinamento, para avaliar se o modelo está evoluindo ou regredindo a cada época passada. Em relação ao grupo de validação, desde o início do treinamento o valor mínimo foi de 85.3%, com uma média de 93.6% e atingindo um valor máximo de 96.3%. O desvio padrão da variação da acurácia no treinamento foi de 2.2%. A presença de uma variação nesta etapa é dada pelo fato de haver aleatoriedade durante o processo.

FIGURA 11 – EVOLUÇÃO DO TREINAMENTO

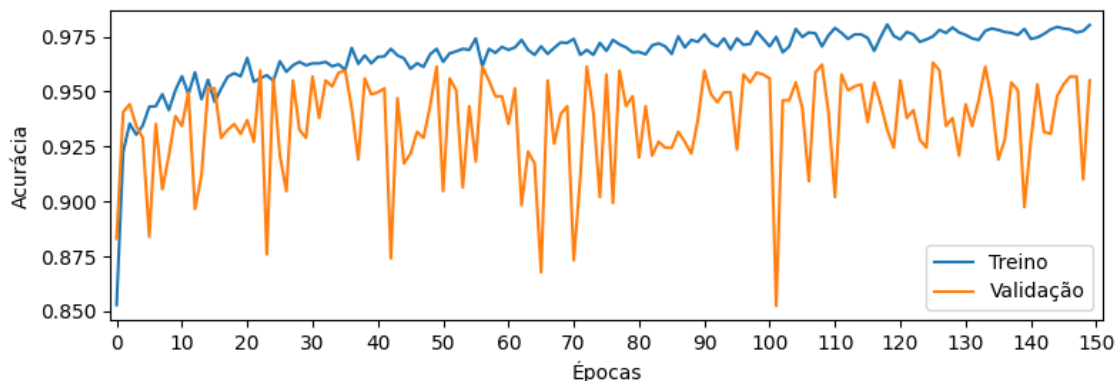

FONTE: O autor (2022).

## 4.3   PERFORMANCE DO MODELO

O modelo foi então exposto ao grupo de teste que não havia sido exposto ainda, cuja composição é de 150 imagens normais e 150 imagens com pneumonia. Das 150 imagens normais, foram identificadas corretamente 147 (VN) enquanto 3 (FP) foram caracterizadas como pneumonia. Das 150 imagens com pneumonia, foram identificadas corretamente 146 (VP), sendo que as outras 4 (FN) foram classificadas como normal (FIGURA 12).

FIGURA 12 – IMAGENS CLASSIFICADAS

FONTE: O autor (2022).

Em termos estatísticos (TABELA 3), a acurácia foi de 97.7%, com sensibilidade de 97.3% e especificidade de 98%. O valor preditivo positivo foi de 98% e o valor preditivo negativo foi de 97.4%. Foi calculado o escore F1, obtendo um valor de 97.7%.

TABELA 3 – PERFORMANCE DO MODELO

| Métrica | Valor |
|---|---|
| Sensibilidade | 97.3% |
| Acurácia | 97.7% |
| Especificidade | 98% |
| Valor preditivo positivo | 98% |
| Valor preditivo negativo | 97.4% |
| Escore F1 | 0.977 |

FONTE: O autor (2022).

## 4.4    VISUALIZAÇÃO DOS RESULTADOS

Na FIGURA 13 estão exemplificadas imagens normais, corretamente identificadas como normais pelo modelo e em seguida sobrepostas com o Grad-CAM. Na FIGURA 14 há imagens com presença de pneumonia, também corretamente identificadas como pneumonia pelo modelo e sobrepostas com o Grad-CAM. As imagens com pneumonia que foram incorretamente identificadas como normais estão expostas na FIGURA 15, enquanto na FIGURA 16 estão expostas as imagens normais incorretamente identificadas como pneumonia.

FIGURA 13 – IMAGENS NORMAIS COM GRAD-CAM

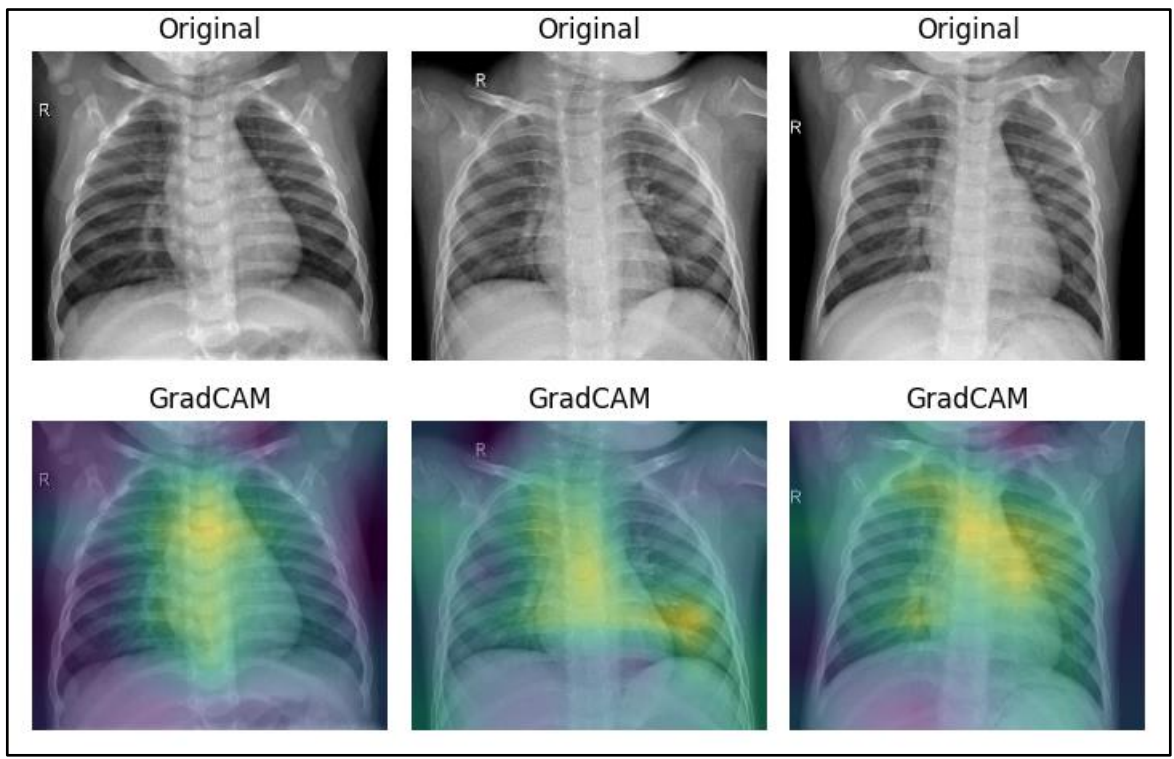

FIGURA 14 – IMAGENS COM PNEUMONIA COM GRAD-CAM

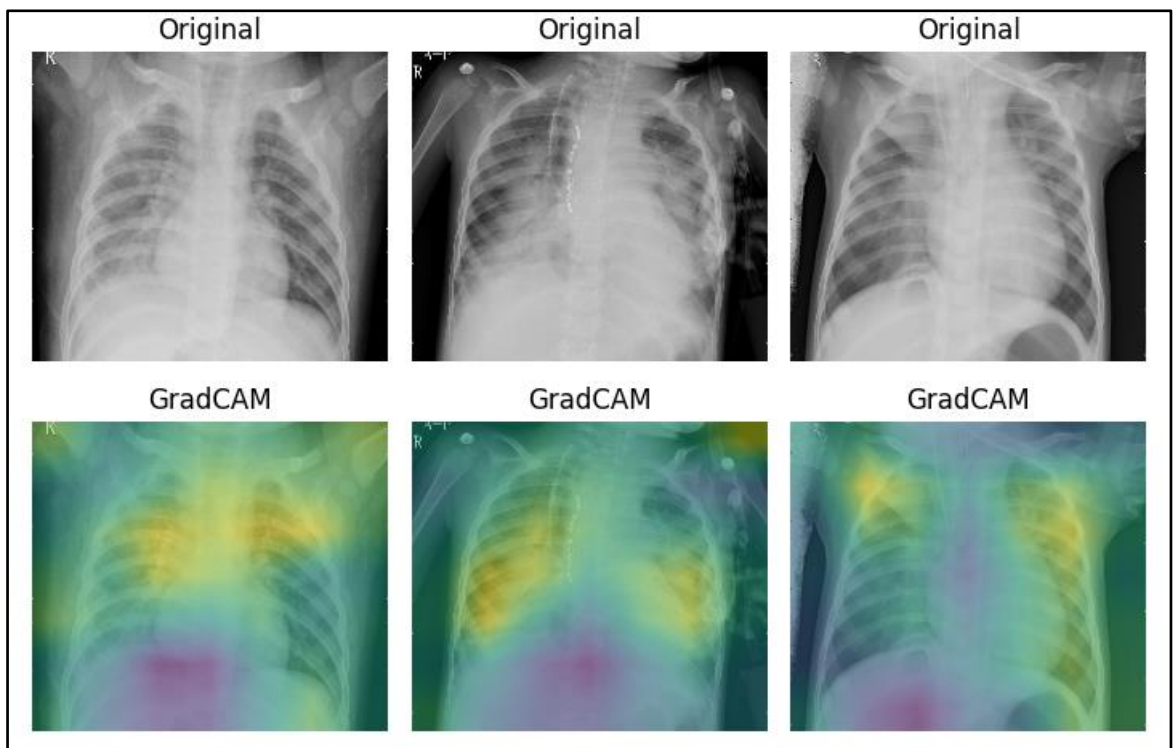

FONTE: O autor (2022).

FIGURA 15 – IMAGENS COM PNEUMONIA INCORRETAMENTE CLASSIFICADAS COMO NORMAIS

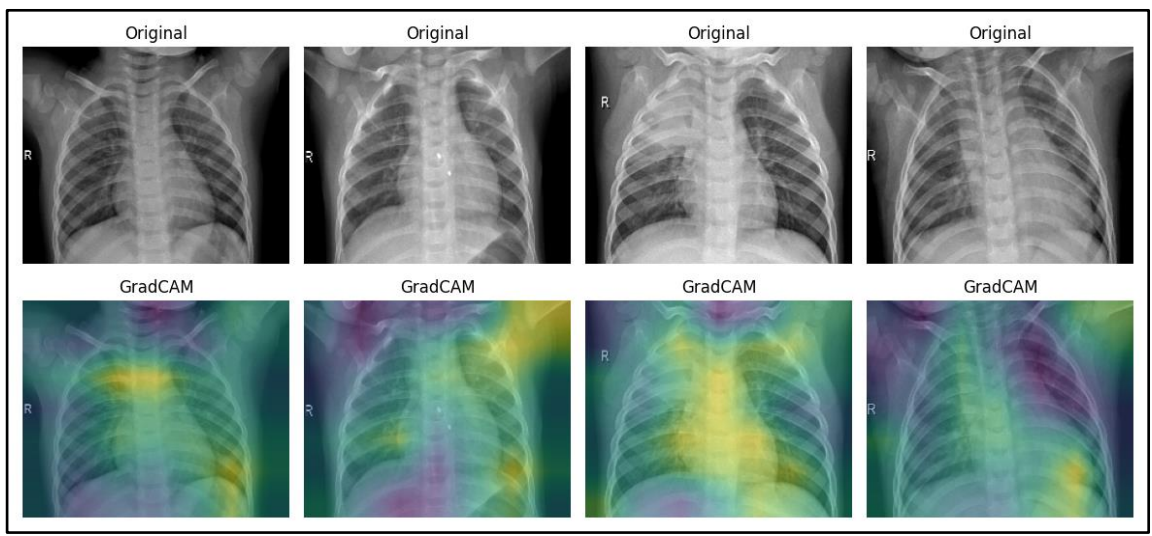

FIGURA 16 – IMAGENS NORMAIS INCORRETAMENTE CLASSIFICADAS COMO PNEUMONIA

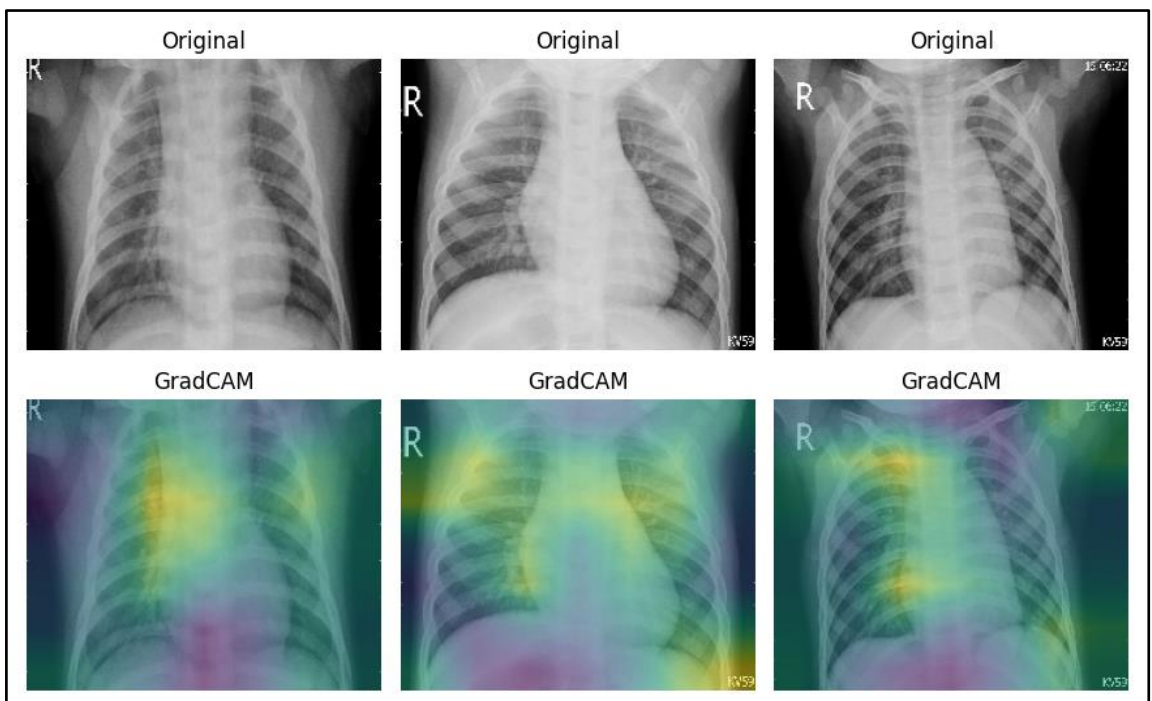

FONTE: O autor (2022).

## 5 DISCUSSÃO

O desenvolvimento de softwares para interpretação de imagens médicas não é algo novo. Esta técnica já foi utilizada por outros autores com diferentes variações de implementações (KERMANY, 2018; KUNDU, 2021; MUHAMMAD, 2021). As estatísticas encontradas na literatura sugerem que o uso de modelos computacionais ajuda a minimizar os erros e pode auxiliar profissionais de saúde com uma segunda opinião.

A implementação utilizada neste trabalho obteve sensibilidade de 97.3% e especificidade de 98%, que, por sua vez, são melhores que os encontrados em alguns trabalhos, a exemplo do trabalho publicado por Kermany *et al.* (2018), cujo modelo atingiu sensibilidade de 93.2% e especificidade de 90.1% para o mesmo banco de imagens. Em ambos os estudos, as estatísticas dos modelos foram geradas apenas pela exposição ao banco de dados utilizado, ainda que dividido sistematicamente. Se os modelos desenvolvidos forem expostos a imagens fora do banco de dados utilizado, o desempenho pode ser diferente. Para confirmar a usabilidade em casos reais, é necessário expor o modelo a imagens fora do banco de dados.

O resultado atingido pelo modelo proposto foi significativo. Tal performance foi atingida devido a, fatores, separação e o pré-processamento das imagens utilizadas foram importantes para tal performance. O uso do *Image Augmentation* proposto é amplamente difundido no contexto de aprendizado de máquina, pois permite que o software receba variações semelhantes ao que a vida real proporciona, como distorções ou perspectivas diferentes. Isto faz com que o modelo esteja melhor preparado para receber imagens diretamente de um operador e evita a necessidade de uma padronização ao interpretar novas imagens.

Apesar do software identificar pneumonia em radiografias, o modelo proposto não se limita a imagens de raio-X. O funcionamento depende apenas dos dados recebidos durante o treinamento, classificando qualquer tipo de informação fornecida, desde que corretamente catalogada – novamente a exemplo de Kermany *et al.* (2018), em que o modelo desenvolvido pelo autor interpretou tanto imagens de tomografias de coerência óptica (OCT) quanto imagens de raio-X. Em outras palavras, o modelo desenvolvido neste trabalho pode ser treinado novamente para outros fins sem ser o de identificar pneumonia, bastaria uma base de dados igualmente catalogada para um novo treinamento de acordo com o que se deseja classificar. Esta versatilidade demonstra que modelos como o proposto possam vir a auxiliar em diversas áreas, inclusive na saúde pública, seja reduzindo custos ou diminuindo a margem de erro.

O processo de desenvolvimento de modelos para aprendizagem de máquina segue um raciocínio sequencial, independentemente da implementação utilizada. Em primeiro momento, é necessário um banco de dados com informações que se desejam classificar, ordenado por rótulos ou classes. Em seguida, deve-se desenvolver um modelo computacional. No caso deste trabalho, o modelo escolhido foi uma rede neural. Após o modelo ser desenvolvido, há o processo de treinamento, o qual consiste em apresentar ao modelo repetidas vezes os dados preparados vindos do banco de dados, de forma que o software crie pesos e medidas para cada parâmetro contido no modelo computacional. Esses pesos são armazenados e utilizados para interpretar dados novos, classificando a informação nova de acordo com os rótulos treinados. Paralelamente ao treinamento, em cada época que se passa, é avaliado a performance do modelo para interpretação do grupo de validação. Essa avaliação permite acompanhar a evolução do modelo ao longo do treinamento, indicando quando estará em sua melhor performance.

Para o treinamento há a necessidade de um alto poder computacional, proporcional ao grau de complexidade do modelo implementado e à quantidade de dados utilizados para treinar. Após o treinamento, o poder computacional exigido para usufruir das capacidades "aprendidas" pelo modelo é ínfimo. Assim, é possível disponibilizar o uso do modelo em dispositivos menos potentes, como smartphones ou diretamente em uma página da web. Isto corrobora com a versatilidade que esta tecnologia pode ser empregada, adicionado também acessibilidade e facilidade de uso como vantagens ao modelo proposto.

Os dados utilizados são suficientes para o treinamento do modelo proposto. Por outro lado, há o que é chamado de "desbalanceamento de classes" (JAPKOWICZ, 2002), que consiste em diferenças na quantidade dos grupos de imagens utilizadas. No banco de dados utilizado há quase 3 vezes a quantidade de imagens do grupo de pneumonia quando comparado ao grupo de imagens normais. Isto cria um viés no treinamento do modelo, o deixando mais propenso a identificar uma imagem como pneumonia do que como normal, o que diminui a qualidade do modelo final, ainda que as estatísticas para o grupo de teste sejam significativas. O ideal seria um banco de imagens em que há uma quantidade semelhante para cada grupo a ser estudado, a fim de evitar que o modelo fique propenso a identificar com mais frequência o grupo que possui mais imagens para treino.

Ao incluir imagens de um mesmo paciente, o modelo é exposto a um viés de seleção. O viés acontece devido a pacientes que têm pneumonia realizarem não só a radiografia inicial, mas também outras radiografias para seguimento, de forma que o mesmo paciente se repita para esta categoria no banco de dados. Considerando que o modelo tem a finalidade de encontrar padrões nos dados recebidos e que várias radiografias se originaram de um só paciente, o padrão encontrado em uma nova radiografia pode ser algo intrínseco ao paciente, como a estrutura do gradil costal, a posição dos membros ou até mesmo o biotipo. Nestes casos, pode-se utilizar o Grad-CAM para avaliação.

Com a visualização das imagens provenientes do Grad-CAM (FIGURA 14 e FIGURA 16), é possível ver que o modelo focou em regiões pulmonares para a identificação da presença de pneumonia, tal qual um humano o faria. Nota-se também que em algumas imagens, mesmo classificando corretamente, o modelo utilizou como área de interesse regiões alheias ao pulmão para sua classificação. No caso da FIGURA 14, regiões fora do tórax foram utilizadas para chegar a tal conclusão, o que não segue uma lógica para o diagnóstico da pneumonia. Situações como estas demonstram a possibilidade de estar identificando características comuns ao paciente e não à opacidade pulmonar em si. Para evitar esta situação, os bancos de dados utilizados poderiam limitar a apenas uma imagem por categoria por paciente, de forma que um paciente não se repita dentro de uma mesma categoria a ser estudada.

Quanto a utilização do modelo em casos reais para diagnosticar um paciente, vale considerar que o software utiliza apenas da imagem para chegar a uma conclusão, enquanto um profissional necessita da clínica do paciente. Sem uma boa clínica não é possível realizar diagnósticos de forma adequada.

# 6 CONCLUSÃO

Através deste trabalho pode-se concluir que é possível desenvolver um software que identifique pneumonia em imagens de raio-X torácico. Ao longo de seu desenvolvimento, foi possível obter imagens em banco de dados públicos, já classificadas de acordo com a presença ou ausência de pneumonia. As imagens coletadas foram analisadas, comparadas e separadas em três grupos com sucesso. Foi possível também desenvolver o modelo de rede neural proposto, sendo treinado com as imagens separadas. Com o modelo já treinado foi possível classificar as imagens propostas. Por fim, foi possível aplicar a técnica Grad-CAM no modelo final com sucesso.

# REFERÊNCIAS